\documentclass[aps,prb,twocolumn,groupedaddress,amsmath,amssymb,showpacs]{revtex4}

\usepackage{graphicx}
\usepackage{dcolumn}
\usepackage{bm}

\begin{document}

\title{Unusual nature of fully-gapped superconductivity in In-doped SnTe}

\author{Mario Novak}
\email{mnovak@sanken.osaka-u.ac.jp}
\author{Satoshi Sasaki}
\author{Markus Kriener}
\thanks{Present address: RIKEN Center for Emergent Matter Science (CEMS), 
Wako 351-0198, Japan} 
\author{Kouji Segawa}
\author{Yoichi Ando}
\email{y_ando@sanken.osaka-u.ac.jp}
\affiliation{Institute of Scientific and Industrial Research, Osaka University, 
Ibaraki, Osaka 567-0047, Japan}

\date{\today}

\begin{abstract}

The superconductor Sn$_{1-x}$In$_x$Te is a doped topological crystalline
insulator and has become important as a candidate topological
superconductor, but its superconducting phase diagram is poorly
understood. By measuring about 50 samples of high-quality, vapor-grown
single crystals, we found that the dependence of the superconducting
transition temperature $T_c$ on the In content $x$ presents a
qualitative change across the critical doping $x_c \simeq$ 3.8\%, at
which a structural phase transition takes place. Intriguingly, in the
ferroelectric rhombohedral phase below the critical doping, $T_c$ is
found to be strongly {\it enhanced} with impurity scattering. It appears
that the nature of electron pairing changes across $x_c$ in
Sn$_{1-x}$In$_x$Te.

\end{abstract}



\maketitle

Recently, superconductors derived from topological insulators are
attracting significant attention, \cite{Fu10, Lee, Hsieh-Fu, Yamakage,
Michaeli, Hor, wray10a, Sasaki, Kriener1, Kriener2, Kriener3, wray11a,
Kirzhner, Georgia, Levy, Bay, Paglione, Sasaki2012, Fang, Sato2013}
because they have a potential to be topological superconductors.
\cite{QiZhang, Ando, Schnyder} Topological superconductors are
characterized by a nontrivial topology of the superconducting wave
functions, \cite{QiZhang, Ando, Schnyder} and they necessarily harbor
gapless surface quasiparticle states that often consist of Majorana
fermions, \cite{Wilczek} an exotic kind of particle that is its own
antiparticle. 

In this context, the superconductor \cite{Bushmarina1986, Erickson2009}
Sn$_{1-x}$In$_x$Te is of interest, because it is derived from SnTe which
is a new type of topological insulator called a topological crystalline
insulator. \cite{Fu2011, Hsieh2012, Tanaka2012} Intriguingly, this
material (with $x \simeq$ 0.045) preserves \cite{Sato2013} the topological
surface state above $T_c$ and, furthermore, was found to present
signatures of surface Andreev bound states in the point contact
spectroscopy. \cite{Sasaki2012} Given that the existence of surface
Andreev bound states is a hallmark of unconventional superconductivity
\cite{Kashiwaya2000} and that the symmetry of the effective Hamiltonian
of Sn$_{1-x}$In$_x$Te implies that an unconventional superconductivity
in this material is bound to be topological, \cite{Sasaki2012}
Sn$_{1-x}$In$_x$Te has become a strong candidate for a topological
superconductor.

Because of the heightened interest in Sn$_{1-x}$In$_x$Te, it is
important to establish its phase diagram with respect to In doping. In
particular, Sn$_{1-x}$In$_x$Te in the low In-doping range is known to
present a ferroelectric transition at low temperature, accompanied by a
structural phase transition from cubic to rhombohedral, \cite{NGS-1983,
Kobayashi1976, Muldawer1973, Brillson1974} and the signatures of surface
Andreev bound states were observed \cite{Sasaki2012} for $x \simeq$
0.045 which is close to this ferroelectric phase. Therefore, it is
useful to clarify the details of the superconducting phase diagram in
the low In-doping range that is only poorly understood.
\cite{Erickson2009} Also, since the superconducting transition
temperature $T_c$ of Sn$_{1-x}$In$_x$Te is unusually high
\cite{Allen1969, Shelankov1987} for its carrier density of
$\sim$10$^{21}$ cm$^{-3}$ and it has been proposed that impurity
scattering might be {\it enhancing} the $T_c$ in this system,
\cite{Martin1997} it would be interesting to see how the $T_c$ is
related to disorder in Sn$_{1-x}$In$_x$Te.

In this Rapid Communication, we report the phase diagram of
Sn$_{1-x}$In$_x$Te for $x$ = 0.018 -- 0.08 based on the measurements of
51 single-crystal samples. It was found that robust, fully-gapped
superconductivity is established for the entire doping range studied,
but the dependence of $T_c$ on $x$ in the ferroelectric rhombohedral
phase is clearly different from that in the cubic phase where $T_c$ vs
$x$ is confirmed to be essentially linear as was previously reported.
\cite{Erickson2009} Our data suggest that the $T_c$ in the rhombohedral
phase is primarily governed by the level of disorder, and surprisingly,
$T_c$ tends to become {\it higher} in more disordered samples. This
tendency seems to exist also in the cubic phase, although there the
$T_c$ is primarily determined by carrier density and the change in $T_c$
with disorder is much weaker. Hence, even though the electron pairing is
expected to be driven by electron-phonon interactions in this material,
\cite{Sasaki2012, Shelankov1987, Martin1997} there is a certain unusual
aspect in the pairing mechanism.

The single crystals of Sn$_{1-x}$In$_x$Te were grown by a
vapor-transport method. A stoichiometric ratio of high purity elements
of Sn (99.99\%), In (99.99\%) and Te (99.999\%) were melted in an
evacuated quartz tube to form a homogeneous polycrystalline source, and
the quartz tube was subsequently transferred to a horizontal three-zone
furnace for the single crystal growth. During the crystal growth, the
source material was kept in a 1.5 K/cm temperature gradient centered at
630 $^\circ$C for 1 week, after which high-quality faceted single
crystals larger than 1 $\times$ 1 mm$^2$ in lateral size are obtained.
The rock-salt structure (space group $Fm\bar{3}m$) of the single
crystals was confirmed by x-ray diffraction analysis at room
temperature. The In content $x$ in the crystals was determined by the
inductively coupled plasma atomic emission spectroscopy (ICP-AES)
analysis, for which the samples were dissolved into 1 M of aqueous
HNO$_{3}$. The measured $x$ values in the vapor-grown samples were
always lower than those of the source materials. We could not achieve
the actual $x$ value of more than 0.08 regardless of the In content in
the source material. Transport measurements were performed using a
six-probe techniques down to 0.3 K in magnetic fields up to $\pm$14T
with the current in the [100] direction. The electrical contacts were
made with Au wires using Ag paint which gave a contact resistance of $<
1$ $\Omega$. Specific heat was measured down to 0.34 K with a
relaxation-time method using a Quantum Design physical property measurement 
system (PPMS-9).

\begin{figure}
\includegraphics[width=5.2cm]{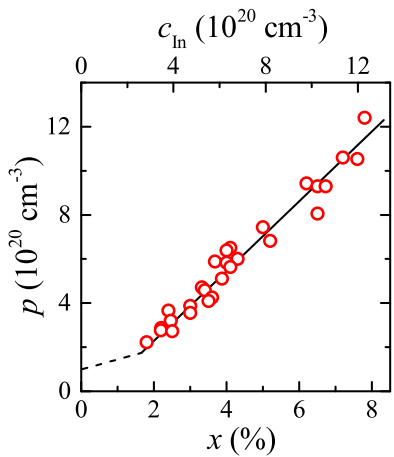}
\caption{Hole density $p$ vs $x$ (lower axis) and $c_{\mathrm{In}}$
(upper axis). The solid line is a least-square fit to the data and the 
dashed line indicates the approximate hole density for $x$ = 0 -- 0.017 
reported in Ref. \onlinecite{Erickson2009}.
}
\end{figure} 

Let us begin by examining the role of In doping. There are two sources
of hole carriers in Sn$_{1-x}$In$_x$Te. The parent compound SnTe is
always Sn deficient and is better written as Sn$_{1-\delta}$Te, where
$\delta$ is usually around $\sim$1\%; \cite{NGS-1983} such Sn vacancies
introduce holes. The second source is the In dopants. The valence of In
in Sn$_{1-x}$In$_x$Te is $+1$,\cite{Erickson2009} and one hole is
introduced per In atom. The ICP-AES analyses allowed us to accurately
determine the actual In content $x$ averaged over the whole volume of
the sample, which also gives the volume density of In atoms,
$c_{\mathrm{In}}$. Alongside this chemical analysis, we determined the
hole density $p$ from the Hall measurements in 29 samples in the
following way: The Hall coefficient $R_{\rm H}$ was extracted from the
slope of the Hall resistivity $\rho_{yx}$ versus the magnetic field $B$,
which is found to be always completely linear. Then, the nominal Hall
carrier density $p_{\mathrm{H}}=1/(e R_{\mathrm{H}})$ is determined at 4
K. The true hole density can be obtained by multiplying $p_{\mathrm{H}}$
with the Hall factor $r$. \cite{Tanaka2012} For SnTe, the Hall factor
has been elucidated to be 0.6, \cite{Houston1964} and hence one obtains
$p$ via $p= 0.6\,p_{\mathrm{H}}$. As shown in Fig. 1, the relation
between $p$ and $c_{\mathrm{In}}$ is linear with a slope close to 1.
This result reassures that the valence of In is $+1$ and that the Hall
factor $r = 0.6$ is valid in Sn$_{1-x}$In$_x$Te. In a previous study,
\cite{Erickson2009} it was found that for $x$ = 0 -- 0.017, $p$ is
primarily determined by Sn vacancies and saturates in the $(1-2)
\times 10^{20}$ cm$^{-3}$ range, which is also indicated in Fig. 1 with
a dashed line. 

\begin{figure}
\includegraphics[width=8cm]{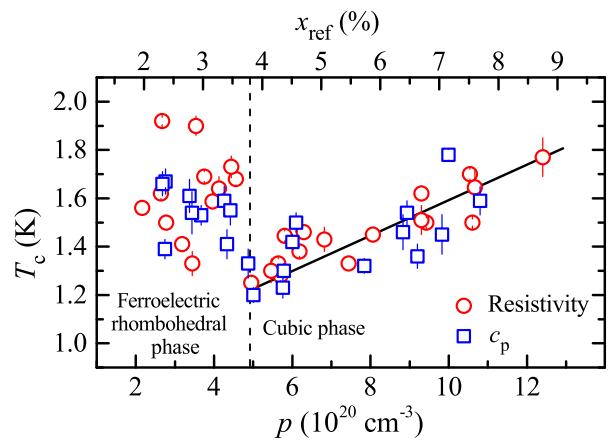}
\caption{$T_c$ vs $p$ plot with the upper horizontal axis showing
$x_{\rm ref}$. Depending on the sample, $T_c$ was measured either with
resistivity (open circles) or with specific heat (open squares).
The vertical dashed line marks $p_c$ to separate the two different structure
phases. The vertical error bar signifies the transition width,
which was defined with the resistivity change from 90\% to 10\% of
$\rho_{\rm 4 K}$, or with the range between the onset and the peak in
the specific-heat jump.}
\end{figure}

Now we focus on the behavior of $T_c$. Figure 2 shows the $T_c$ vs $p$
plot for all the samples measured in this work. The $T_c$'s of 29
samples were measured with resistivity, and those of 22 samples
were measured with specific heat. In both cases, $T_c$ is determined
from the mid-point of the transition, and the error bar signifies the
transition width. The data are primarily shown against $p$, because we
found that $T_c$ shows better systematics against $p$ rather than
against $x$. \cite{note_x} In Fig. 2, using the upper horizontal axis,
we also show $x_{\rm ref}$, which is calculated from $p$ using the
linear relation found in Fig. 1, to give reference to the doping level
per formula unit in this plot.

\begin{figure*}
\includegraphics[width=16cm]{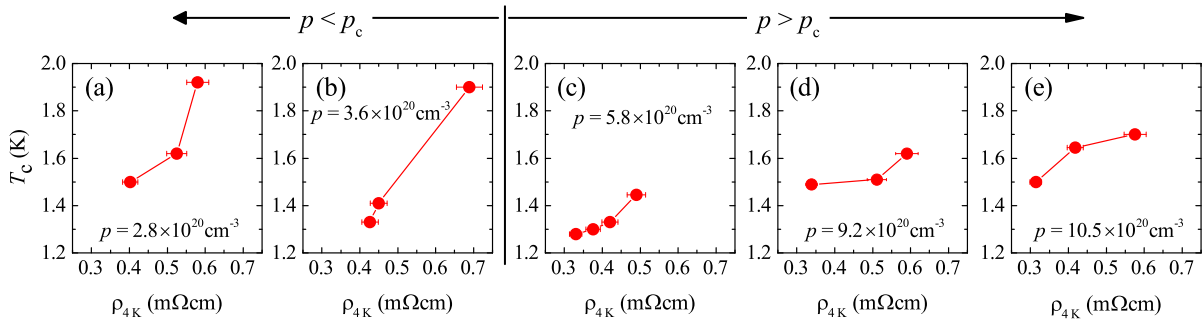}
\caption{
Dependencies of $T_c$ on $\rho_{\mathrm{4 K}}$ for five $p$ values 
(in units of $10^{20}$ cm$^{-3}$), 2.8, 3.6, 5.8, 9.2, and 10.5.
The first two are in the ferroelectric rhombohedral phase ($p < p_c$),
while the last three are in the cubic phase ($p > p_c$).}
\end{figure*}
  
The lower limit of $p$ where we found superconductivity is 2.2 $\times$
10$^{20}$ cm$^{-3}$, which corresponds to $x_{\rm ref} \simeq$ 0.019 and
is consistent with the previous report. \cite{Erickson2009} However,
while it was reported in Ref. \onlinecite{Erickson2009} that
$T_{\mathrm{c}}$ linearly increases with increasing $x$ once
superconductivity with $T_c >$ 0.3 K is induced in samples with $x
\gtrsim 0.02$, we found that such a trend only exists for $p$ above a
critical value $p_c \simeq 4.8 \times 10^{20}$ cm$^{-3}$, which
corresponds to $x_{\rm ref} \simeq$ 0.038. Strikingly, this $p_c$ is
essentially the same as the critical hole density $p_c^{\rm FE}$ above
which the ferroelectric rhombohedral phase
disappears.\cite{Erickson2009, noteFE} Below this critical $p_c$, we
found that $T_c$ shows virtually no correlation with $p$ and spreads
rather widely between 1.3 and 1.9 K. It is interesting to note that the
highest $T_c$ of 1.9 K in this study was observed in samples at low
doping, $p \simeq 2.7 \times 10^{20}$ and $3.5 \times 10^{20}$
cm$^{-3}$, and it exceeds the $T_c$ of maximally doped samples. On the
other hand, the lowest $T_c \approx 1.2$ K was only observed in samples
near $p_c$ on the cubic-structure side.

To understand this puzzling behavior, we show in Fig. 3 the plots of
$T_c$ versus the residual resistivity at 4 K, $\rho_{\mathrm{4 K}}$, for
several $p$ values at which three or more samples were measured. Note
that $\rho_{\mathrm{4 K}}$ gives a measure of the strength of impurity
scattering in the sample. One can see a clear trend that $T_c$ is {\it
enhanced} as $\rho_{\mathrm{4 K}}$ becomes larger. In particular, the
highest $T_c$ of 1.9 K was observed in samples with a relatively large
$\rho_{\mathrm{4 K}}$ of 0.6 -- 0.7 m$\Omega$cm, which is very unusual.
This unusual trend is most obvious in samples with $p < p_c$ and it is
weaker for $p > p_c$, but it seems that the same trend still exists at
large $p$. Altogether, the results shown in Figs. 2 and 3 strongly
suggest that $T_c$ is mainly determined by the unusual enhancement due
to impurity scattering in the ferroelectric rhombohedral phase, while in
the cubic phase it is determined primarily by $p$ but is still affected
by the impurity scattering in an unusual manner.

It is well known that in unconventional superconductors, the pairing gap
is strongly suppressed by impurity scattering. \cite{Balian, Larkin,
Meckenzie, MaenoRMP, Maeno12} In conventional BCS superconductors, on
the other hand, $T_c$ is known to be insensitive to weak disorder.
\cite{Anderson} The present observation is at odds with both cases, and
hence is very peculiar. Nevertheless, it was argued by Martin and
Phillips \cite{Martin1997} that nonmagnetic impurities in
Sn$_{1-x}$In$_x$Te and Pb$_{1-x}$Tl$_x$Te could enhance $T_c$ due to
their dual role of reducing the on-site Coulomb repulsion and enhancing
the density of states at the Fermi energy (in this theory, the
negative-$U$ mechanism \cite{Varma} is not involved). Therefore, our
observation is not without theoretical justification. Apparently, more
microscopic studies to elucidate the pairing mechanism in
Sn$_{1-x}$In$_x$Te are strongly called for.

\begin{figure}[b]
\includegraphics[width=8.5cm]{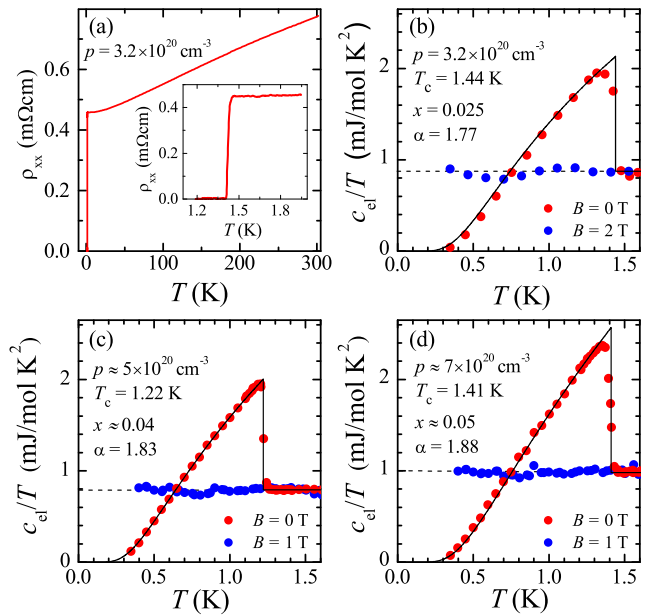}
\caption{(a) $\rho_{xx}(T)$ of a sample in the ferroelectric rhombohedral
phase ($p < p_c$), $p = 3.2 \times 10^{20}$
cm$^{-3}$ ($x$ = 0.025); the inset show the sharp superconducting 
transition at $T_c$ = 1.42 K. 
(b) $c_{\mathrm{el}}/T$ vs $T$ of the
same sample measured down to 0.34 K in 0 T (red circles) and in 2 T
(blue circles). 
(c), (d) $c_{\mathrm{el}}/T$ vs $T$ plots for samples in the cubic phase
($p > p_c$), $p \simeq 5 \times 10^{20}$ cm$^{-3}$ ($x \simeq$ 0.04,
$T_c$ = 1.22 K) and $p \simeq 7 \times 10^{20}$ cm$^{-3}$ ($x \simeq$
0.05, $T_c$ = 1.41 K); those samples were only characterized with 
specific heat. In (b)--(d), solid lines show the modified BCS
fits to the experimental data with tunable $\alpha$, and dashed lines
indicate the values of $\gamma_{\mathrm{n}}$. }
\end{figure}

Given that the behavior of $T_c$ for $p < p_c$ is very unusual and that
it was previously suggested \cite{Erickson2009} that bulk
superconductivity may not be established in this regime, it is useful to
investigate the specific-heat anomaly associated with the
superconductivity in this phase. Figure 4(a) shows the temperature
dependence of the resistivity $\rho_{xx}$ of a sample in the
ferroelectric rhombohedral phase having $p = 3.2 \times 10^{20}$
cm$^{-3}$ ($x$ = 0.025), which showed $\rho_{\mathrm{4 K}}$ = 0.46
m$\Omega$cm and $T_c$ = 1.42 K with a narrow transition width of 22 mK.
The characteristic kink in the $\rho_{xx}(T)$ behavior associated with
the structural phase transition
\cite{Erickson2009} is not clearly seen in this sample,
probably because the resistivity at the transition is already dominated
by impurity scattering rather than by phonon scattering. 
The specific-heat data of the same sample is shown in Fig. 4(b) in terms
of $c_{\mathrm{el}}/T$ vs $T$. Here, $c_{\mathrm{el}}$ is the
electronic specific heat obtained after subtracting the phonon
contribution from the total specific heat. \cite{note_cp} The
normal-state electronic specific-heat coefficient $\gamma_{\mathrm{n}}$
= 0.87 mJ/mol K$^2$ is indicated by the dashed line. One notices that
$c_{\mathrm{el}}/T$ in 0 T approaches zero at low temperature, which
gives evidence that this sample is 100\% superconducting and that a
fully-gapped superconductivity can be robustly established even in the
ferroelectric rhombohedral phase.

The $c_{\mathrm{el}}/T$ data in 0 T was fitted with the modified BCS
theory \cite{Padamsee} which allows variation of the coupling constant
$\alpha$ ($\equiv \Delta_0/T_c$, with $\Delta_0$ the superconducting gap
at 0 K). The fitting shown in Fig. 4(b) is made so that it correctly
reproduces the jump near $T_c$; \cite{m*} the resulting $\alpha$, 1.77,
is essentially the same as the weak-coupling value of 1.76. The data at
lower temperature deviate only slightly from the fitting curve,
suggesting that the BCS theory gives a reasonably good description of
the specific-heat behavior. 
 
To complement the above specific-heat data for the rhombohedral phase,
similar data measured on samples in the cubic phase ($p > p_c$) are
shown in Figs. 4(c) and 4(d) for $p \simeq 5 \times 10^{20}$ cm$^{-3}$
($x \simeq$ 0.04, $T_c$ = 1.22 K) and $p \simeq 7 \times 10^{20}$
cm$^{-3}$ ($x \simeq$ 0.05, $T_c$ = 1.41 K), respectively. The fittings
using the modified BCS theory are again made so that they correctly
reproduce the jump near $T_c$; the obtained $\alpha$ values, 1.83 and
1.88, are larger than that for $x$ = 0.025, but otherwise the data are
reasonably well described by the modified BCS theory.

Now we discuss the implications of our data. Between the ferroelectric
rhombohedral phase and the cubic phase, the difference in the doping
dependence of $T_c$ seems to suggest that the pairing mechanism is
somewhat different. Such an inference is corroborated by the fact that
the unusual enhancement of $T_c$ with impurity scattering is prominently
observed in the rhombohedral phase, while this effect is much weaker in
the cubic phase. In this regard, it is useful to note that the
Martin-Phillips theory \cite{Martin1997} for the enhancement of $T_c$
due to impurity scattering is within the framework of the BCS theory,
and therefore the superconductivity in the rhombohedral phase is not
necessarily unconventional. \cite{note_gamma} 

We would like to mention that we have measured one sample with $T_c$ =
1.9 K in the rhombohedral phase with the same point-contact technique
used in Refs. \onlinecite{Sasaki} and \onlinecite{Sasaki2012} and found
that the spectra present a two-peak structure that is akin to the
conventional Andreev reflection spectra described by the
Blonder-Tinkham-Klapwijk theory.\cite{BTK} On the other hand, we have
consistently observed a single zero-bias conductance peak in seven
samples so far measured with $T_c \simeq$ 1.2 K. We therefore speculate
that in Sn$_{1-x}$In$_x$Te the even- and odd-parity pairing states may
be competing and, since the odd-parity state is expected to be
suppressed with nonmagnetic impurities,\cite{Balian, Larkin, Meckenzie,
MaenoRMP, Maeno12} the conventional even-parity state wins in samples
with high $T_c$'s where impurity scattering is strong. If this is indeed
the case, the odd-parity state is realized only in the lowest $T_c$
samples where the impurity scattering is the weakest. 

In the ferroelectric rhombohedral phase, there is no Brillouin-zone
folding and the Fermi surface volume is expected to be unchanged from that
in the cubic phase. Nevertheless, establishment of the ferroelectric phase 
is accompanied with generations of ferroelectric domains and charge
polarizations, which may have something to do with the present
observation. In particular, ferroelectric domain boundaries
are expected to be pinned by structural defects in crystals, and hence
more disordered samples would contain higher densities of ferroelectric
domain boundaries to cause strong electron scattering.  

It is prudent to note that all the specific-heat data are reasonably
well described by the BCS theory throughout the doping range, which
implies that the gap magnitude is always nearly isotropic. However, this
does not necessarily contradict the possible realization of odd-parity
pairing in low $T_c$ samples suggested by the point-contact spectroscopy
made on samples with $T_c \simeq$ 1.2 K, because the specific-heat
behavior is expected to be essentially the same between the isotropic
even-parity pairing and the fully-gapped odd-parity pairing.
\cite{Hashimoto} In this context, the coupling constant $\alpha$ to
phenomenologically explain the data is found to be slightly {\it larger}
in the $T_c$ = 1.22 K sample [Fig. 4(c)] compared to the $\alpha$ in the
$T_c$ = 1.44 K sample [Fig. 4(b)], which is at odds with the natural
expectation that stronger coupling leads to higher $T_c$; this seems to
support the idea that the pairing mechanism is different between the two
phases.

In summary, we have elucidated the superconducting phase diagram of
Sn$_{1-x}$In$_x$Te as a function of In doping for $x < 8\% $ and found
that the nature of electron pairing is possibly different between the
ferroelectric rhombohedral phase ($x \lesssim$ 3.8\%) and the cubic
phase ($x \gtrsim$ 3.8\%). In particular, in the former phase the $T_c$
was found to be strongly {\it enhanced} with impurity scattering, while
such an effect is weaker in the cubic phase where $T_c$ seems to be
primarily governed by carrier density. The unusual role of impurity
scattering suggests that conventional even-parity pairing is likely to
be realized in higher $T_c$ samples and unconventional superconductivity
may only be found in cleaner, lower $T_c$ samples.

\begin{acknowledgments}
We thank T. Ueyama and A. A. Taskin for their help in crystal growths, and 
L. Fu and P. Phillips for helpful discussions.
This work was supported by JSPS 
(KAKENHI 24740237, 24540320, and 25220708), MEXT (Innovative Area 
``Topological Quantum Phenomena" KAKENHI), and AFOSR (AOARD 124038).
M.N. acknowledges financial support from Croatian Science Foundation 
(Grant No. O-1025-2012).
\end{acknowledgments}


\begin{thebibliography}{50}

\bibitem{Fu10}
L. Fu and E. Berg, 
Phys. Rev. Lett. \textbf{105}, 097001 (2010).

\bibitem{Lee}
L. Hao and T. K. Lee, 
Phys. Rev. B {\bf 83}, 134516 (2011). 

\bibitem{Hsieh-Fu}
T. H. Hsieh and L. Fu, 
Phys. Rev. Lett. {\bf 108}, 107005 (2012).

\bibitem{Yamakage}
A. Yamakage, K. Yada, M. Sato, and Y. Tanaka, 
Phys. Rev. B {\bf 85}, 180509(R) (2012).

\bibitem{Michaeli}
K. Michaeli and L. Fu, 
Phys. Rev. Lett. {\bf 109}, 187003 (2012).

\bibitem{Hor}
Y. S. Hor, A. J. Williams, J. G. Checkelsky, P. Roushan, J. Seo, Q. Xu, 
H. W. Zandbergen, A. Yazdani, N. P. Ong, and R. J. Cava, 
Phys. Rev. Lett. \textbf{104}, 057001 (2010). 

\bibitem{wray10a}
L. A. Wray, S.-Y. Xu, Y.~Xia, Y. S. Hor, D.~Qian, A. V. Fedorov, H.~Lin,
A.~Bansil, R. J. Cava, and M. Z. Hasan,
Nature Physics {\bf 6}, 855 (2010).

\bibitem{Sasaki}
S. Sasaki, M. Kriener, K. Segawa, K. Yada, Y. Tanaka, M. Sato, and Y. Ando, 
Phys. Rev. Lett. \textbf{107}, 217001 (2011).

\bibitem{Kriener1}
M. Kriener, K. Segawa, Z. Ren, S. Sasaki, and Y. Ando, 
Phys. Rev. Lett. \textbf{106}, 127004 (2011).

\bibitem{Kriener2}
M. Kriener, K. Segawa, Z. Ren, S. Sasaki, S. Wada, S. Kuwabata, and Y. Ando,
Phys. Rev. B \textbf{84}, 054513 (2011).

\bibitem{Kriener3}
M. Kriener, K. Segawa, S. Sasaki, and Y. Ando:
Phys. Rev. B {\bf 86} (2012) 180505.

\bibitem{wray11a}
L. A. Wray, S.~Xu, Y.~Xia, D.~Qian, A. V. Fedorov, H.~Lin, A.~Bansil, L.~Fu, 
Y.S. Hor, R. J. Cava, and M. Z. Hasan,
Phys.\ Rev.\ B {\bf 83}, 224516 (2011).

\bibitem{Kirzhner}
T. Kirzhner, E. Lahoud, K. B. Chaska, Z. Salman, and A. Kanigel,
Phys. Rev. B {\bf 86}, 064517 (2012).

\bibitem{Georgia}
X. Chen, C. Huan, Y. S. Hor, C. A. R. Sa de Melo, and Z. Jiang,
arXiv:1210.6054.

\bibitem{Levy}
N. Levy, T. Zhang, J. Ha, F. Sharifi, A. A. Talin, Y. Kuk, 
and J. A. Stroscio, 
Phys. Rev. Lett. {\bf 110}, 117001 (2013).

\bibitem{Bay}
T. V. Bay, T. Naka, Y. K. Huang, H. Luigjes, M. S. Golden, and A. de Visser,
Phys. Rev. Lett. {\bf 108}, 057001 (2012).

\bibitem{Paglione}
K. Kirshenbaum, P. S. Syers, A. P. Hope, N. P. Butch, J. R. Jeffries, 
S. T. Weir, J. J. Hamlin, M. B. Maple, Y. K. Vohra, and J. Paglione, 
Phys. Rev. Lett {\bf 111}, 087001 (2013).

\bibitem{Sasaki2012}
S. Sasaki, Z. Ren, A. A. Taskin, K. Segawa, L. Fu, and Y. Ando, 
Phys. Rev. Lett. {\bf 109}, 217004 (2012).

\bibitem{Fang}
L. Fang, C.C. Stoumpos, Y. Jia, A. Glatz, D.Y. Chung, H. Claus, U. Welp,
W.K. Kwok, and M.G. Kanatzidis, arXiv:1307.0260.

\bibitem{Sato2013} 
T. Sato, Y. Tanaka, K. Nakayama, S. Souma, T. Takahashi, S. Sasaki, Z. Ren, 
A. A. Taskin, K. Segawa, and Y. Ando, 
Phys. Rev. Lett \textbf{110}, 206804 (2013). 


\bibitem{QiZhang}
X.-L. Qi and S.-C. Zhang, 
Rev. Mod. Phys. \textbf{83}, 1057-1110 (2011).

\bibitem{Ando}
Y. Ando, 
J. Phys. Soc. Jpn. {\bf 82}, 102001 (2013).

\bibitem{Schnyder} 
A. P. Schnyder, S. Ryu, A. Furusaki, and A. W. W. Ludwig, 
Phys. Rev. B \textbf{78}, 195125 (2008).

\bibitem{Wilczek}
F. Wilczek, Nature Phys. {\bf 5}, 614 (2009).

\bibitem{Bushmarina1986} 
G. S. Bushmarina, I. A. Drabkin, V. V. Kompaniets, R. V. Parfenev, 
D. V. Shamshur, and M. A. Shakhov, 
Sov. Phys. Solid State \textbf{28}, 612 (1986).

\bibitem{Erickson2009} 
A. S. Erickson, J.-H. Chu, M. F. Toney, T. H. Geballe, and I. R. Fisher, 
Phys. Rev. B \textbf{79}, 024520 (2009).

\bibitem{Fu2011}
L. Fu, Phys. Rev. Lett. {\bf 106}, 106802 (2011).

\bibitem{Hsieh2012} 
T. H. Hsieh, H. Lin, J. Liu, W. Duan, A. Bansil, and L. Fu, 
Nature Commun. \textbf{3}, 982 (2012).

\bibitem{Tanaka2012} Y. Tanaka, Z. Ren, T. Sato, K. Nakayama, S. Souma, 
T. Takahashi, K. Segawa, and Y. Ando, 
Nature Phys. \textbf{8}, 800  (2012).

\bibitem{Kashiwaya2000} 
S. Kashiwaya and Y. Tanaka, 
Rep. Prog. Phys. \textbf{63}, 1641 (2000).

\bibitem{NGS-1983} 
R. Dornhaus, G. Nimtz, and B. Schlicht, 
\textit{Narrow-Gap Semiconductors}, Springer Tracts in Modern Physics 
(Springer, New York, 1983), Vol. 98. 

\bibitem{Kobayashi1976} 
K. L. I. Kobayashi, Y. Kato, Y. Katayama, and K. F. Komatsubara, 
Phys. Rev. Lett. \textbf{37}, 772 (1976). 

\bibitem{Muldawer1973} 
L. Muldawer, J. Nonmet. \textbf{1}. 177 (1973).

\bibitem{Brillson1974} 
L. J. Brillson,  E. Burstein, and L. Muldawer, 
Phys. Rev. B \textbf{9}, 1547 (1974). 

\bibitem{Allen1969} 
P. B. Allen and M. L. Cohen, Phys. Rev. \textbf{177}, 704 (1969).

\bibitem{Shelankov1987} 
A. L. Shelankov, Solid State Commun. \textbf{62}, 327 (1987).

\bibitem{Martin1997} 
I. Martin and P. Phillips, Phys. Rev. B \textbf{56}, 14650 (1997).

\bibitem{Houston1964} 
B. B. Houston, R. S. Allgaier, J. Babiskin, and P. G. Siebenmann, 
Bull. Amer. Phys. Soc. \textbf{9}, 60 (1964).

\bibitem{note_x}
For those samples in which the Hall coefficient was not measured, the
$p$ value was assigned based on the $x$ value determined from ICP-AES.

\bibitem{noteFE}
In Ref. \onlinecite{Erickson2009}, the highest value of $x$
where the ferroelectric phase was observed was 0.034, which suggests
the critical doping $x_c^{\rm FE}$ should be around 0.04.

\bibitem{Balian}
R.~Balian and N. R. Werthamer,
Phys.\ Rev. {\bf 131}, 1553 (1963).

\bibitem{Larkin}
A. I. Larkin, JETP Lett. {\bf 2}, 130 (1965).

\bibitem{Meckenzie} 
A. P. Mackenzie, R. K. W. Haselwimmer, A. W. Tyler, G. G. Lonzarich, Y. Mori, 
S. Nishizaki, and Y. Maeno, Phys. Rev. Lett. \textbf{80}, 161 (1998).

\bibitem{MaenoRMP}
A. P. Mackenzie and Y.~Maeno,
Rev.\ Mod.\ Phys. {\bf 75}, 657 (2003).

\bibitem{Maeno12}
Y.~Maeno, S.~Kittaka, T.~Nomura, S.~Yonezawa, and K.~Ishida,
J.\ Phys.\ Soc.\ Jpn. {\bf 81}, 011009 (2012).

\bibitem{Anderson} 
P. W. Anderson, J. Phys. Chem. Solids \textbf{11}, 26 (1959).

\bibitem{Varma} 
C. M. Varma, Phys. Rev. Lett. \textbf{61}, 2713 (1988). 

\bibitem{note_cp}
The total specific heat $c_p$ in the normal state after suppressing the
superconductivity in 1 or 2 T was fitted with the Debye formula $c_p =
\gamma_{\rm n}T + A_3 T^3 + A_5 T^5$, where $A_3$ and $A_5$ are the
coefficients of the phononic contribution. The fitting parameters for
the samples shown in Figs. 4(b), 4(c), and 4(d) are: 
($\gamma_{\rm n}$ [mJ/mol K$^2$], $A_3$ [mJ/mol K$^4$], $A_5$ [mJ/mol K$^6$]) 
= (0.874, 0.655, 0.016), (0.786, 0.704, 0.007), and (0.982, 0.578, 0.014), 
respectively.

\bibitem{Padamsee}
H.~Padamsee, J.E. Neighbor, and C.A. Shiffman,
J.\ Low Temp.\ Phys. {\bf 12}, 387 (1973).

\bibitem{m*}
From the data in Figs. 4(a) and 4(b), we can estimate the effective mass
via $m^* = (3\hbar^2 \gamma_{\mathrm{n}}) / (V_{\mathrm{mol}} k_B^2 k_F)
= 3.4\,m_e$ ($V_{\mathrm{mol}}$ = 38 cm$^3$/mol is the molar volume and
$m_e$ is the free electron mass) by approximating the Fermi wave number
$k_F$ = $[3\pi^2 (p/4)]^{1/3}$ = 1.3 nm$^{-1}$, for which spherical Fermi
surfaces located at the four $L$ points are assumed. The mean free path
$\ell = \hbar k_F/(\rho_{\rm 4 K} p e^2)$ = 3.9 nm is much shorter than
the coherence length $\xi_{0} = \hbar v_F/(\pi \alpha k_B T_c)$ = 38 nm.

\bibitem{note_gamma}
The $\gamma_{\rm n}$ value for $x$ = 0.025 [Fig. 4(b)] is larger than
that for $x \simeq$ 0.04 [Fig. 4(c)] despite the lower carrier density,
which is consistent with the notion (Ref. \onlinecite{Martin1997}) that an
enhancement of the density of states is responsible for higher $T_c$. 

\bibitem{BTK}
G. E. Blonder, M. Tinkham, and T. M. Klapwijk, 
Phys. Rev. B {\bf 25}, 4515 (1982).

\bibitem{Hashimoto}
T. Hashimoto, K. Yada, A. Yamakage, M. Sato, Y. Tanaka, 
J. Phys. Soc. Jpn. \textbf{82}, 044704 (2013).


\end{thebibliography}

\end{document}